\documentclass[twocolumn]{aastex631}

\def\simgr{\,\hbox{\hbox{$ > $}\kern -0.8em \lower 1.0ex\hbox{$\sim$}}\,}
\def\simle{\,\hbox{\hbox{$ < $}\kern -0.8em \lower 1.0ex\hbox{$\sim$}}\,}
\newcommand{\nswift}{Neil Gehrels Swift Observatory BAT}
\newcommand{\xmm}{XMM-Newton}
\newcommand{\obj}{Swift~J0503.7$-$2819}
\def\pola{1RXS~J083842.1$-$282723}
\def\polb{IGR~J19552+0043}
\def\polc{SDSS~J084617.11+245344.1}
\def\pal{RX~J0524+42}

\shortauthors{Halpern}
\shorttitle{Swift~J0503.7$-$2819: Asynchronous Polar}

\begin{document}
\title{Swift~J0503.7$-$2819: A Short-Period Asynchronous Polar or Stream-Fed Intermediate Polar}

\author[0000-0003-4814-2377]{J. P. Halpern}
\affiliation{Department of Astronomy and Columbia Astrophysics Laboratory, Columbia University, 550 West 120th Street, New York, NY 10027-6601, USA; jph1@columbia.edu}

\begin{abstract}
    We analyze a 7.4~hr \xmm\ light curve of the cataclysmic variable \obj, previously classified using optical periods as an intermediate polar (IP) with an orbital period of 0.0567~days.  A photometric signal at 975~s, previously suggested to be the spin period, is not present in X-rays and is readily understood as a quasi-periodic oscillation.   The X-ray light curve instead shows clear behavior of a highly asynchronous polar (AP) or stream-fed IP.  It can be described by either of two scenarios: one which switches between one-pole and two-pole accretion, and another in which accretion alternates fully between two poles.  The spin periods in these two models are 0.0455 days and 0.0505 days, respectively.  The spin frequency $\omega$ is thus either 24\% faster or 12\% faster than the orbital frequency $\Omega$, and the corresponding beat period between spin and orbit is 0.231~days or 0.462~days.  Brief absorption events seen in light curve are spaced in a way that may favor the longer spin and beat periods. These periods are confirmed and refined using data from the Transiting Exoplanet Survey Satellite (TESS) and the Asteroid Terrestrial-impact Last Alert System (ATLAS).  The short beat cycle of \obj\ makes it well-suited to resolving this common dilemma, which amounts to deciding whether the main signal in the power spectrum is $\omega$ or $2\omega-\Omega$.
\end{abstract}

\section{Introduction\label{sec:intro}}

Cataclysmic variables (CVs) are accreting binaries in which a dwarf
star donates mass to a white dwarf (WD) via Roche-lobe overflow.
X-ray surveys preferentially select CVs in which the magnetic
field of the WD is strong enough to truncate the accretion disk
at a magnetospheric boundary, or even prevent a disk from forming
entirely.  In these systems, an accretion stream is channeled onto
the magnetic pole(s), where thermal plasma heated by a shock
in the column just above the surface of the WD radiates X-rays.
In polars (AM Her stars), the magnetic field is strong enough to
channel matter directly from the companion, and the WD rotation
is locked to the binary orbit.  Polars are also characterized
by optical circular polarization, and optical/IR humps in their
spectra from cyclotron radiation in the strong magnetic field. 
Intermediate polars (IPs, or DQ Her stars) have weaker magnetic
fields and a truncated accretion disk.  The spin period of the
WD in an IP is detected as a coherent oscillation in X-ray or
optical emission from a rotating hot spot, at a shorter 
period than the orbital period of the binary.

In addition, there is a small group of ``asynchronous polars'' (APs),
stream-fed systems in which the spin and orbit periods differ by $\le2\%$.
The four original members of this group are V1500~Cyg, BY~Cam,
V1432~Aql, and CD~Ind (\citealt{cam99}, and references
therein).  V1500 Cyg had a nova explosion in 1975 \citep{sto88},
and because it is generally observed that APs
are evolving toward synchronism on a short time-scale of hundreds
of years, it is believed that nova eruptions
perturb the spin of the WD to create the asynchronism \citep{sch91}.
Recently, three more members were discovered, \pola\ \citep{hal17,rea17},
\polb\ \citep{tov17}, and \polc\ \citep{lit22}, but they are asynchronous
by $3-4\%$.  Finally, the peculiar
object \pal, also known as Paloma \citep{sch07,jos16,lit22},
has been added to this group.  Its spin period is 13\% shorter
than its orbital period.  The more extreme APs may be better described
as either pre-polars approaching synchronism for the first time,
or stream-fed or diskless IPs if their periods are in stable equilibrium.

The subject of this Paper is the hard X-ray selected CV \obj, which we \citep{hal15} identified from the \nswift\ survey.  Its Gaia-CRF3 position is R.A.=$05^{\rm h}03^{\rm m}49^{\rm s}\!.260$, decl.=$-28^{\circ}23^{\prime}07^{\prime\prime}\!.96$ \citep{bro21}, which is referenced for proper motion to epoch 2016.0.  Proper motion components are ($\mu_{\alpha}{\rm cos}\,\delta, \mu_{\delta})=(+5.02\pm0.07, +21.87\pm0.08)$~mas~yr$^{-1}$.   Its parallax is $1.142\pm0.086$~mas, corresponding to a distance of 837~(794--897)~pc using geometric  priors \citep{bai21}.  Its X-ray luminosity is $3.6\times10^{32}$ erg~s$^{-1}$ in the 0.2--12~keV band from \xmm\ \citep{web20}, and $2.4\times10^{32}$ erg~s$^{-1}$ in the 14--195~keV Swift BAT survey band\footnote{https://swift.gsfc.nasa.gov/results/bs157mon/}.

In \citet{hal15} we measured an orbital period of 0.0567~days in \obj,
both from radial velocity spectroscopy
and time-series photometry.  We interpreted another photometric signal at
975~s as a spin period, which would make \obj\
a rare example of an IP below the 2--3~hr (orbital) period gap in CVs.
Here, we present an analysis of an
archival X-ray observation (Section~\ref{sec:xray}) that does not show
a 975~s period.  Instead, it reveals two possible values
for the spin period, both of these close the orbital period.
As a result, classification
of \obj\ as an AP or a nearly synchronous stream-fed IP is preferred.
The optical evidence is revisited in view of this new information
in Section~\ref{sec:opt}.  More precise values for the periods
are found in archival survey photometry.  The difficulty of identifying the correct
spin period, and the arguments for and against the presence of an
accretion disk, are discussed in Section~\ref{sec:disc}.
Conclusions are presented in Section~\ref{sec:conc}.

\section{X-ray Light Curve and Power Spectrum\label{sec:xray}}

\obj\ was observed by \xmm\ on 2018 March~7 (ObsID 0801780301) for 7.4~hr.
We used the processed event files from the two EPIC MOS cameras and the EPIC pn cameras to create X-ray light curves.
Fast-mode data were taken by the Optical Monitor in the $V$ filter.  Although these are too faint to be useful for timing analysis, the mean magnitude
of $V=18.1$ is consistent with historical values (see Section~\ref{sec:opt}).

\begin{figure*}[ht]
\centerline{
\includegraphics[angle=270.,width=1.0\linewidth]{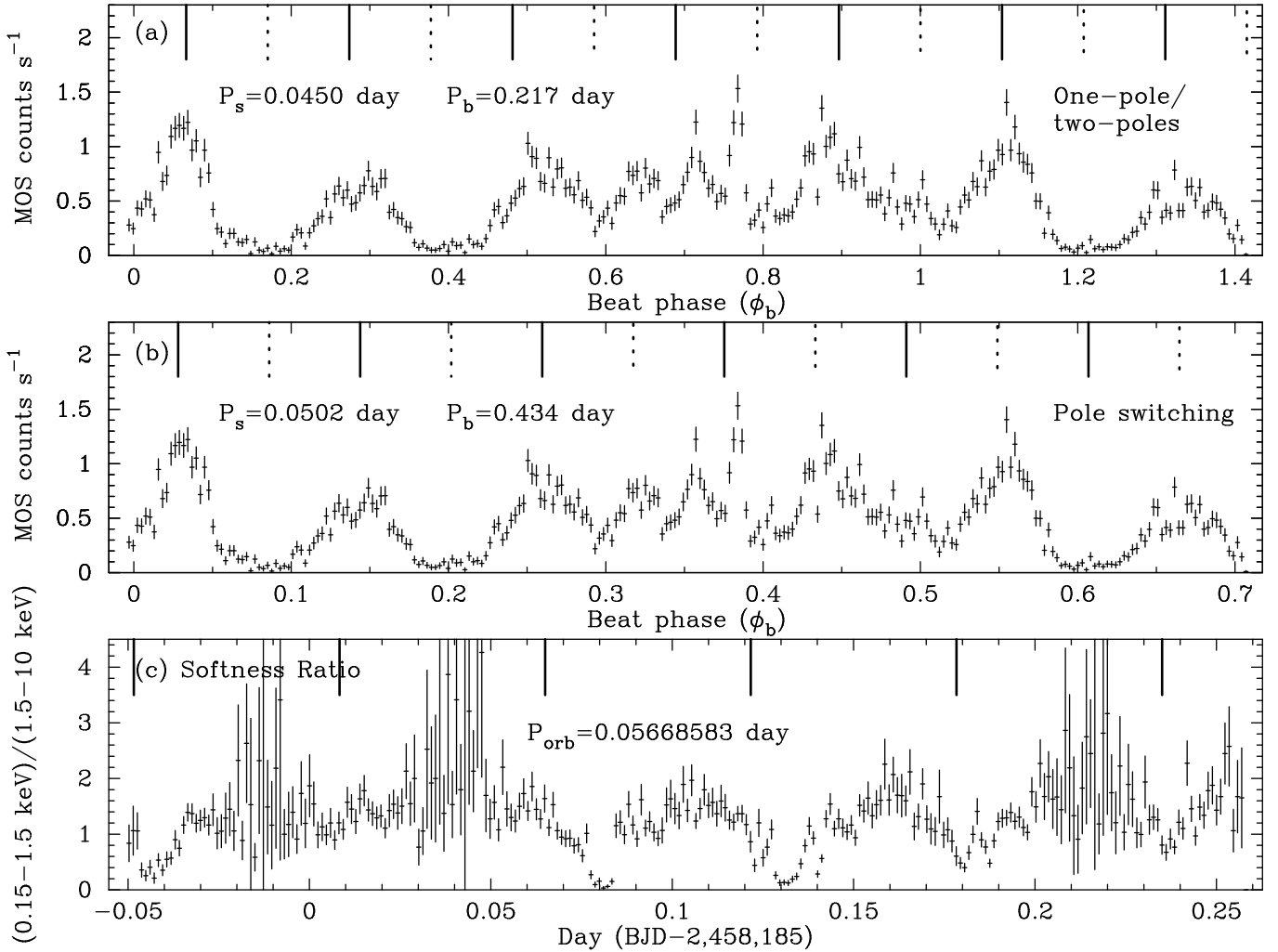}
}
\caption
{\xmm\ EPIC MOS light curve (0.2--10~keV) in 100~s bins, marked with the meridian crossing times of the two accreting poles using the alternative model spin periods ($P_{\rm s}$) from Table~\ref{tab:periods}, and the orbital period.  (a) Solid lines mark a dominant pole that accretes continuously. The secondary pole (dotted lines) only sometimes accretes.  An entire beat period ($P_{\rm b}$) is contained within the observation. Phase 0 of the beat cycle is arbitrary. (b) The pole-switching model requires a longer spin period, and neither pole accretes continuously.  The beat period is twice as long as above, and is not completely covered by the observation. (c) Ratio of counts in the 0.15--1.5~keV band to the 1.5--10~keV band.  The EPIC pn data have been included to improve the statistics.  The solid lines mark orbital phase 0 defined as the epoch of blue-to-red crossing of the optical emission-line radial velocity (see Section~\ref{sec:atlas})}.
\label{fig:mos}
\end{figure*}

Figure~\ref{fig:mos} shows the light curve from the combined MOS cameras, which
have 1610~s longer exposure time than the pn. The softness ratio
(0.15--1.5~keV)/(1.5--10~keV) is also displayed, using both
the pn and the MOS to improve the statistics.   The main feature of the light
curve is one that is common in polars: periodic flat dips which
are each too long to be an eclipse by the secondary star or its accretion
stream.  They are not due to photoelectric absorption, as their softness is
maintained throughout.  Rather, they are caused by the WD self-occulting a hot
spot near its surface.  Since the emission lasts
for more than half of the rotation, the spot must be in the ``upper''
hemisphere of the WD, i.e., closest to the observer.  However, there are
additional peaks (or weakening of dips) over half of the observation that
are not accounted for by a single hot spot, and may indicate emission from
another location.  These will be investigated with the help of
a power-spectrum analysis.
In addition, there are three or four shorter dips in the softness
ratio, indicating a drastic drop in the soft (0.15--1.5~keV) X-rays due to
photoelectric absorption.

\begin{figure*}
\centerline{
\includegraphics[angle=270.,width=1.\linewidth]{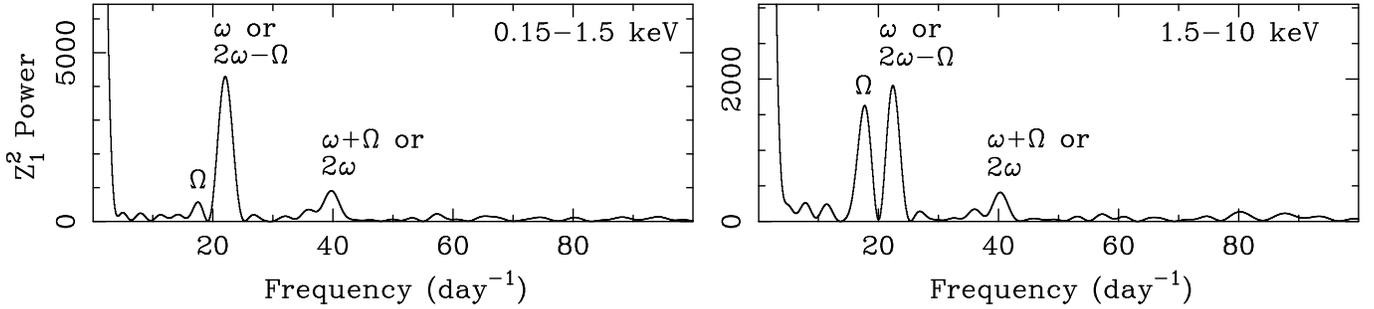}
}
\caption
{Periodograms of the combined pn and MOS data for the soft (0.15--1.5 keV) and hard (1.5--10 keV) X-rays.  A periodogram of the total energy range was used to measure the frequencies listed in Table~\ref{tab:periods}. Alternative identifications of two of the three marked peaks are indicated.  These alternatives define models 1 and 2.  The identification of the orbital frequency $\Omega$ is unambiguous and is used for both models in Table~\ref{tab:periods}.}
\label{fig:power}
\end{figure*}

We calculated $Z_1^2$ periodograms from the barycentered photon arrival
times from the combined pn and MOS data in their overlapping time spans.
Shown in Figure~\ref{fig:power}, the power
spectra have three main peaks, but with different relative strengths in
soft and hard X-rays.  We tentatively adopt the periods from a
combined 0.15--10~keV
power spectrum and list them in Table~\ref{tab:periods}.  Note that
the uncertainties listed in Table~\ref{tab:periods} are purely statistical,
and are smaller than possible systematic effects due to the limited span of the
observation with respect to the various periods found.
Therefore, the precision implied by the number of digits used here
should not be taken literally. More precise values found optically
will be given in Section~\ref{sec:opt}.

The only X-ray signal that is familiar from our optical study \citep{hal15}
is the lowest frequency one, at 17.62~day$^{-1}$ (0.0567~day). It is
identical within errors to the 0.0567~day optical orbital period.
But this is not the period between the X-ray peaks, which is
better represented by the peak at 22.23~cycles~day$^{-1}$ (0.0450~day).
This is one reason why we maintain that the optical period is in fact
the orbital period.
In order to try to explain the detailed behavior of the X-ray light curve,
we next propose two possible scenarios, each involving contributions
from two emitting poles.  In the first model, only one of
the poles is accreting continuously.  In the second model, accretion
switches completely between poles and the spin period has a different
value from the first model.

\begin{deluxetable}{cccc}
\label{tab:periods}
\tablecolumns{4} 
\tablecaption{X-ray Observed and Inferred Frequencies}
\tablehead{
  \colhead{Frequency} & \colhead{Period} & \colhead{Model 1}
 & \colhead{Model 2} \\
\colhead{(day$^{-1}$)} & \colhead{(day)} & \colhead{(One-pole/two-poles)} &
\colhead{(Pole switching)}
}
\startdata
17.62(4)   &  0.0567   &  $\Omega$         &  $\Omega$          \\
22.23(3)   &  0.0450   &  $\omega$         &  $2\omega-\Omega$  \\
39.97(7)   &  0.0250   &  $\omega+\Omega$  &  $2\omega$         \\
\hline
19.93      &  0.0502   &    \dots          & $\omega$           \\
4.61       &  0.217   &  $\omega-\Omega$  &   \dots            \\
2.31       &  0.434   &    \dots          & $\omega-\Omega$    \\
\enddata
\tablecomments{Frequencies above the line are the three observed in the periodogram (Figure~\ref{fig:power}); those below the line are inferred.}
\end{deluxetable}

\subsection{Model 1: One-pole/two-poles \label{sec:one}}

Assuming that the 22.23 cycles~day$^{-1}$ (0.0450~day) peak is the spin period,
the times when each of two hot spots, separated by $180^{\circ}$ of longitude,
cross the meridian, are marked in Figure~\ref{fig:mos}a.
The dominant spot emits continuously, but extra emission
partly fills in between four consecutive pulses.  If this is emission
from a second pole in antiphase to the first, it means that the accretion
stream switches from feeding one pole exclusively, to feeding both poles
for approximately half of the beat cycle between the spin and orbit.
The beat frequency $\Omega_{\rm b}=\omega-\Omega$ corresponds to a period
of 0.217~day, which is shorter than the length of the observation.
By the sixth and seventh observed pulse from the dominant pole,
the pattern has started to repeat and again only the dominant pole
is accreting.

There are two or three sharp dips in the softness ratio beginning at day 0.082
(Figure~\ref{fig:mos}c).  This coincides with the time that the second pole
starts accreting, and it occurs on its meridian crossing.  It is possible
that this represents absorption by the newly fed stream accreting onto the
second pole, enabled by its particular geometry. e.g., if it is in the
lower hemisphere.  The absorption at this
phase then weakens for each successive rotation.  Finally the second pole
stops accreting, and only the dominant pole emits.  However, the spacing
between the absorption dips is slightly longer than the spin period, and it is not
clear what the role is of a possible dip at the very beginning of the
observation, on day $-0.045$.

The effects of the absorption on the power spectra are dramatic.
In soft X-rays, emission from the second pole is suppressed,
while the dominant pole continues to accrete.  With little
evidence in soft X-rays that there is another pole,
the beating effect is weak and the
orbital frequency is suppressed in the power spectrum.
In the hard X-rays, the feeding of the second pole flattens
the spin modulation and creates power at the beat frequency.  This in
turn creates sidebands of the spin signal at
$\omega \pm \Omega_{\rm b} = \Omega$ and $2\omega-\Omega$.
The $\Omega$ (lower) sideband is prominent in hard X-rays, while the upper
sideband is present, but very weak.  It is not clear what produces this
asymmetry.  Perhaps it is the short observation not being commensurate with
the beat period.  There are also an uncertain number of narrow flares
that complicate any simple picture.

The third-highest peak in the power spectrum, at a frequency of
39.97~day$^{-1}$ (0.0250~day), is consistent with being
the sum of the first two, namely $\omega+\Omega$
But is not clear what would produce it.  It is natural to
get $|\omega-\Omega|$ if a single pole is accreting.
But to get $\omega+\Omega$ would require either retrograde
spin, or pole-switching with asymmetric poles \citep{wyn92}.

\subsection{Model 2: Pole switching\label{sec:two}}

A different description of the power spectrum is commonly considered for APs
that show evidence for complete pole-switching, that is, when only one pole
accretes at a time.  
This was first described by \citet{wyn92} in the context of IPs, but is 
applied frequently to APs (see \citealt{wan20}, \citealt{lit22},
and references therein). If the poles are diametrically opposed but
both visible, then the light curve consists of alternating segments,
each modulated at the spin period, but phase-shifted by $180^{\circ}$.
The effect on the power spectrum is to drastically weaken the spin signal
$\omega$, replacing it with the two sidebands at $\omega \pm \Omega_{\rm b} =
\Omega$ and $2\omega-\Omega$.  This differs from model~1, in which $\omega$
has a different value and remains strong in the power spectrum.

In model~2, we identify the two strongest peaks in the power spectrum
of the hard X-rays with these two sidebands, $\Omega=17.62$~day$^{-1}$ and
$2\omega-\Omega=22.23$~day$^{-1}$.  The identification of $\Omega$
is therefore unchanged, while the now unseen spin frequency is inferred to
be 19.93~day$^{-1}$ (0.0502~days), midway between the two peaks.
The beat period is $(\omega-\Omega)^{-1}=0.434$~days, twice the length of
the beat period in model 1.  Therefore, the observation
did not cover the entire beat cycle in this model.

There is an interval when both poles accrete, between day 0.07 and 0.19
in Figure~\ref{fig:mos} when the switch is slowly taking place.  The third peak
in the power spectrum, now identified with $2\omega$, would come from this
interval.  Before and after that there is clearly only one pole visible
because the flux goes almost to zero between pulses.  A different pole
accretes at the end of the observation than at the beginning.  In an
antipodal geometry in which the flux from each pole goes to zero during its
occultation by the WD, the sum of the observer's viewing angle $i$ from the spin
axis and the magnetic dipole inclination $m$, must be $>90^{\circ}$.
In addition, for the length of the occultation of the two spots to be nearly
the same, as seems to be the case, the angle $i$ should be large, but not so
close to $90^{\circ}$ as to allow eclipses of the white dwarf by the secondary.
No such solid-body eclipses are seen.

In the idealized model with no photoelectric absorption, the sidebands
$\Omega$ and $2\omega-\Omega$ have the same strength
\citep{wyn92,wan20}, which is approximately the case for the hard (1.5--10~keV)
X-rays in Figure~\ref{fig:power}.  For the soft (0.15--1.5~keV) X-rays,
the $\Omega$ sideband is reduced by photoelectric absorption suppressing
some of the emission from the second pole, as indicated by the dips in the
softness ratio during the pole-switching interval.

What about the location of the absorption dips in this model?  Here they
come midway between the peaks of the accreting poles, that is, when the
poles are at quadrature phase in their rotation and both are accreting.
If we envision a switching accretion stream stretching out to reach a new
pole, then it does seem possible that the stream will be viewed along
its length at the limb of the WD, and occult that pole as it
rises.  The spacing between the dips is more compatible with the
spin period of model 2 than model 1, although this is somewhat ambiguous
because there is substructure in the dips.  But the spacing is certainly
incompatible with the orbital period.  Interestingly, the possible additional
dip at day $-0.045$ can be understood as coming 2.5 rotation periods
before the next dip at day $0.082$.
This could mean that it is the last occurring absorption event from the
other pole.  If so, these dips favor model 2, since they are evidence for the
existence of the other half of the hypothesized 0.434~day beat cycle.

\begin{figure*}
\centerline{
  \includegraphics[angle=0.,width=0.87\linewidth]{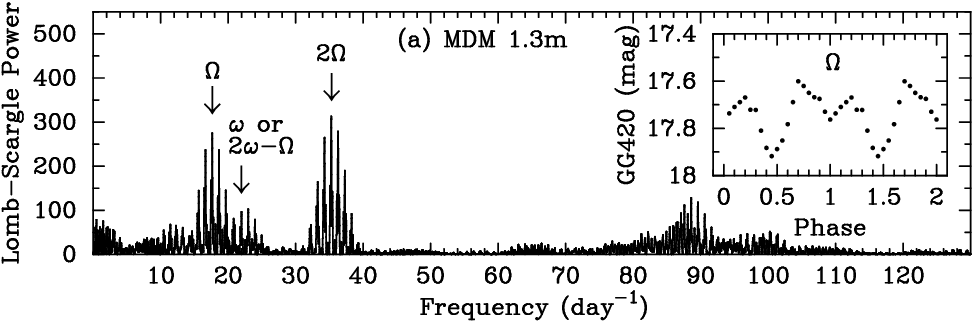}
}
\centerline{
    \includegraphics[angle=0.,width=0.87\linewidth]{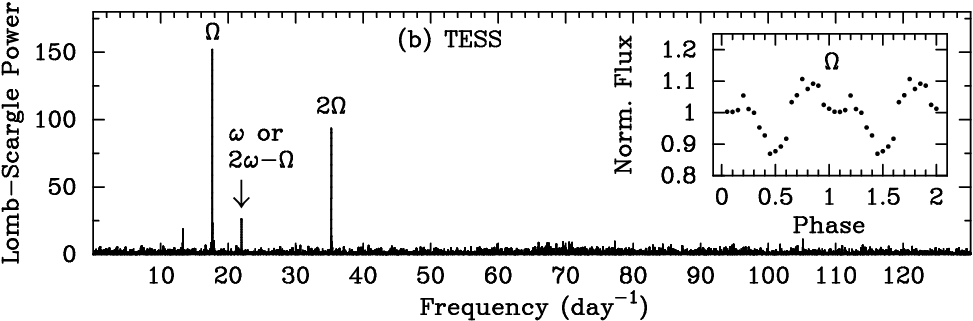}
}
\centerline{
    \includegraphics[angle=0.,width=0.87\linewidth]{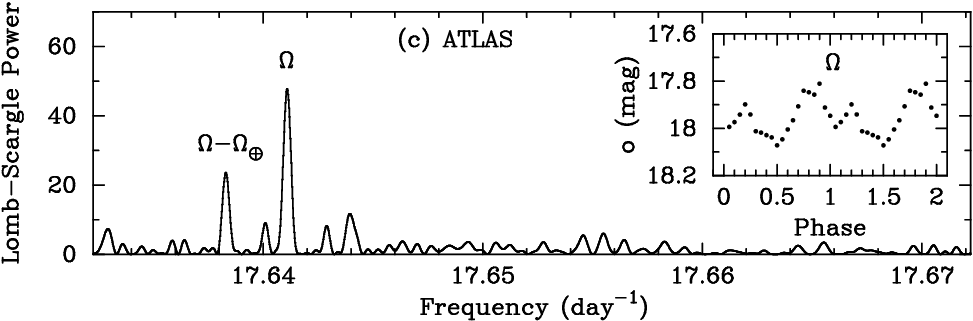}
}
\caption
    {(a) The periodogram of the optical time-series photometry of \obj\
      from \citet{hal15}.  The orbital frequency $\Omega$ is consistent
      with the emission-line radial velocity value in the same study.
      Power centered around 88~cycles~day$^{-1}$ is now attributed to a QPO.
      (b)  The periodogram from TESS finds the same signals as MDM and
      \xmm\, but with higher precision.
      (c) The periodogram of the ATLAS $o$ filter photometry from 2015--2022
      shows the orbital frequency with yet higher precision.
      The second-highest peak is
      a 1~yr alias of $\Omega$.  In all panels the insets are the light
      curves folded
      on the ATLAS orbital period and repeated once. Phase 0 is the epoch of
      blue-to-red crossing of the emission-line radial velocity
      (see Section~\ref{sec:atlas}).}
    \label{fig:opt}
\end{figure*}

\section{Optical Evidence\label{sec:opt}}

\subsection{MDM Observatory (2013 December--2014 January)}

\citet{hal15} reported a spectroscopic orbital period of 0.05662(9)~day from
emission lines, primarily \ion{He}{1} and \ion{He}{2}, and a consistent
period of 0.05667(5)~day from time-series optical photometry in a broad-band
GG420 filter. However, the strongest photometric signal was at the harmonic,
0.028345(12)~day.  Figure~\ref{fig:opt}a shows the periodogram of the time
series from that study. The inset shows the light curve folded on the orbital
period.  In addition to the orbital frequency $\Omega$
and its harmonic $2\Omega$, a possible detection of a signal corresponding
to either the spin frequency $\omega$ or $2\omega-\Omega$ from the
X-rays is marked.  Like the other signals, it is aliased by the daily sampling.
A peak at 21.98~day$^{-1}$ is the one that is closest
to the X-ray value of 22.23~day$^{-1}$.

Power around 88 cycles~day$^{-1}$, previously attributed
to a spin period, is apparently a quasi-periodic oscillation (QPO).
This is clear because (1) it is not
present in X-rays, (2) its power is more broadly distributed in frequency than
the other signals in Figure~\ref{fig:opt}a, and (3) its origin can be deduced
from the optical light curves in \citet{hal15}, where short series of intense
but intermittent flares with the appropriate spacing are seen.  These may
be of the same origin as fast flares in the X-ray light curve near day 0.11
in Figure~\ref{fig:mos}.  Such $\approx1000$~s QPOs are common in CVs, and
\citet{war04} relates them to the rotation period of vertically extended
structures in the outer disk both reflecting and obscuring the central
source.

\subsection{TESS (2020 November 20--December 16)\label{sec:tess}}

A 120~s cadence light curve of \obj\ was obtained by the Transiting Exoplanet
Survey Satellite (TESS; \citealt{ric15}), spanning 26 days starting on
2020 November 20.  We downloaded the processed
light curve and used it to calculate the periodogram in Figure~\ref{fig:opt}b.
It shows the same features as the MDM and \xmm\ data, but with much greater
precision, and free of aliases.  Notably, the $\omega$ or $2\omega-\Omega$
signal is clearly detected at 21.971~day$^{-1}$, as is a weaker signal at
13.308~day$^{-1}$.  The latter is either $2\Omega-\omega$ or $3\Omega-2\omega$.
It was also possibly present in the MDM power spectrum among several
aliases.  The light curve folded on $\Omega$ looks much the same as the
earlier MDM one, but there is less power in the harmonic than in the
fundamental.

We consider that the frequencies measured by TESS (and also ATLAS, see below)
are consistent with the ones from \xmm\ and should supplant them, considering
the systematic effects that must plague the short X-ray observation.
Table~\ref{tab:opt} lists the optically determined values and their
identifications with the model frequencies.  Subsequent discussion
adopts these values.

\subsection{ATLAS (2015 October--2022 January)\label{sec:atlas}}

We downloaded photometry of \obj\ from the Asteroid Terrestrial-impact Last
Alert System (ATLAS; \citealt{ton18}) forced photometry server, which is
useful for confirming and extending the record of the optical behavior
of \obj.  After filtering out points with uncertainties $>0.2$~mag,
there are 1148 observations in the ``orange'' $o$ filter (560--820 nm) spanning
2300 days between 2015 October and 2022 January.  The mean magnitude is
fairly constant during this entire period at $o=17.97$, which is similar to
the GG420 magnitude calibrated approximately as $o=17.75$ in the
MDM light curve, and $V=18.1$ in the \xmm\ OM.   A periodogram of the ATLAS
data (Figure~\ref{fig:opt}c) shows the same orbital period as the MDM data,
but it is much more precise because of the long time span.  The period is
0.05668583(14)~days. The folded light curve looks much the same as the ones
from MDM and TESS.
The weaker signals involving the spin (not shown in Figure~\ref{fig:opt}c)
are also listed in Table~\ref{tab:opt}.

The high precision of the ATLAS period enables us to establish the phase of
all of the folded light curves in Figure~\ref{fig:opt} with respect to the MDM
radial velocity curve, even though the MDM observations were obtained
$\approx 620$~days before the start of ATLAS.  The light curves are folded
on the ATLAS period, with
phase~0 corresponding to the blue-to-red crossing of the emission-line
radial velocity, at BJD 2,456,683.6274(9) \citep{hal15}.  The same ephemeris
is also applied to the 2018 \xmm\ light curve in Figure~\ref{fig:mos}c.

\begin{deluxetable}{llcc}
\label{tab:opt}
\tablecolumns{4} 
\tablewidth{0pt} 
\tablecaption{Optical Observed and Inferred Frequencies}
\tablehead{
 \colhead{Frequency} & \colhead{Period} & \colhead{Model 1}
 & \colhead{Model 2} \\
 \colhead{(day$^{-1}$)} & \colhead{(day)} & & 
}
\startdata
& \hfill TESS & & \\
\hline
 13.3085(55)   & 0.075140(31)   &  $2\Omega-\omega$ & $3\Omega-2\omega$\\
 17.6395(20)   & 0.056691(6)    &  $\Omega$     &  $\Omega$ \\
 21.9717(57)   & 0.045513(12)   &  $\omega$     &  $2\omega-\Omega$ \\
 35.2820(27)   & 0.028343(2)    &  $2\Omega$    &  $2\Omega$      \\
\hline
& \hfill ATLAS & & \\
\hline
 13.30867(7)   & 0.0751390(4)   &  $2\Omega-\omega$ & $3\Omega-2\omega$\\
 17.641093(44) & 0.05668583(14) &  $\Omega$     &  $\Omega$ \\
 21.97355(12)  & 0.04550927(25)  &  $\omega$     &  $2\omega-\Omega$ \\
 35.282186(80) & 0.028342915(64)&  $2\Omega$    &  $2\Omega$      \\
 \hline
&  \hfill Inferred  &  \\
 \hline
 19.80732(6)   &  0.0504864(17)  &    \dots          & $\omega$     \\
  4.33242(8)   &  0.230818(4)   &  $\omega-\Omega$  &   \dots    \\
  2.16621(4)   &  0.461636(9)   &    \dots          & $\omega-\Omega$ \\
\enddata
\end{deluxetable}

\subsection{Interpretation}

Now that we are considering an interpretation as an AP, 
there is ambiguity in interpreting an optical period that
potentially arises from emission in an accretion stream.
The stream leaves the secondary star and ends on the WD, which are
rotating at different frequencies. 
If there is pole switching, the stream
has to break and reform at a different location, causing phase jumps.
Furthermore, the emission-line source may not be at the location of the
optical continuum that modulates
the broad-band photometry, the latter being cyclotron radiation
from low in the accretion column, while the
optical lines are emitted higher up.  Unlike in a synchronous polar,
the accretion stream in an AP
can reach a long way around the WD (see references in \citealt{lit15}).
The emission lines can also have
contributions from the ballistic accretion stream and the heated face
of the secondary.  Therefore, it is not obvious whether the optical
emission should be tied to the rotating
frame of the WD or the orbital frame of the secondary.

In the case of \obj, it is likely that the 0.0567~day period
is the orbital period because it disagrees with either of the
two candidates for the X-ray pulse period, which is most naturally the spin.
We also checked specifically for evidence of the X-ray spin period
candidates in the radial velocity data, including if they had been
suppressed by pole switching, but found no indication that pole switching
had affected the radial velocities, and therefore no evidence for the
spin period itself.

If the emission-line velocities are orbital, they are too broad to be
coming from the heated face of the secondary, while \citet{hal15} state
that they do not show the asymmetric, shifting wings typical of a polar.
Therefore, it is possible that they come from an accretion disk.
The radial velocity amplitude of $189\pm20$~km$^{-1}$ is larger
than that of most IPs, but smaller than that of most polars.
There are also two potential sources of optical continuum
light: the disk, and a stream that manages to pass over
the disk and accrete directly onto a magnetic pole.  A stream is
necessary in order to make the WD poles aware of the orbital period,
evident in the X-ray power spectrum, information which they 
would not have in the case of a conventional disk-fed IP.

Note also that optical emission lines from accretion
disks generally do not represent the orbital velocity
of the compact object, which in this case should be
$\approx60$~km~s$^{-1}$ assuming a $0.8\,M_{\sun}$ WD with
a $0.1\,M_{\sun}$ secondary star.  Rather, their velocities
are due to asymmetric features in the fast-rotating disk.  Therefore,
they are an unreliable indicator of orbital phase even while they
reveal the orbital period.  This complicates any interpretation of
the folded optical light curve in Figure~\ref{fig:opt}
that attempts to relate it to spectroscopic phase.

Without knowing the absolute orbital phase,
the double-peaked shape of the folded optical light curve
in Figure~\ref{fig:opt} is open to more than one interpretation,
but none is highly satisfactory.  The first could involve primary
and secondary eclipses, but the main broad dip is too wide to be
an eclipse.
Second, it could represent a combination of ellipsoidal modulation
and heating of the inner face of the secondary by the X-ray source.
The heating contributes most at the fundamental frequency, while
the ellipsoidal modulation is responsible for the harmonic.
However, the strong optical flickering and, more importantly, absence of
photospheric absorption lines in the spectrum, do not allow for a
major contribution of the secondary's photosphere to the optical light,
while in this interpretation the photosphere must contribute at least 30\%.

Third, the double peaks could represent emission from a more elongated,
optically thick accretion funnel or stream coming off the secondary,
which has a larger projected area as seen from the side.  This is often
invoked for accretion columns in general, but ones that are attached to
the spinning frame, not the orbiting frame that we are restricted to here.
This is essentially a more extreme version of the second proposal, but removed
from the stellar photosphere and also not attached to the WD.
Finally, it may just be easiest for a disk to reveal the orbital
signature in optical light, as it directly taps the majority of
the available gravitational potential energy, and may have persistent
asymmetry driven by accretion from the secondary.

\section{Discussion\label{sec:disc}}

The basic properties of nearly all APs are difficult to determine because
it is not certain which periods represent the spin and the orbit.
It is most helpful if at least the orbital period can be identified
unambiguously by a radial velocity curve of photospheric absorption
lines on the secondary, but these are often not detectable, including in \obj.
Instead, we have a fairly secure but not ironclad case
for $\Omega$ as the optical emission-line and photometric period.
A very precise period derived from 6.3~yr of ATLAS
  photometry allows the optical light curves, radial velocity curve,
  and X-ray light curve to be aligned in phase.  How this
  corresponds to the absolute phase of the binary orbit remains
  an open question.

The spin period can in principle be measured from a segment
of the X-ray light curve between pole switches as we did for
\pola\ \citep{hal17}, an analysis that benefits from strong and
regular pulse shapes.  However, even in such clear cases there
is the complication that the footpoint of the accretion column
is not expected to be fixed on the surface of a WD in an AP \citep{gec97}.
Seen from a fixed point on the WD, the secondary orbits in the
retrograde direction if $\omega>\Omega$.  As the secondary transits over
a pole, the accreting matter is captured and threaded onto different
field lines.   The result is seen clearly as a spin-phase drift during every
beat cycle \citep{lit19,lit22}, although it is not clear if the model of
\citet{gec97} correctly describes the result in all APs \citep{lit19}.
Such an effect would be difficult to recognize and quantify in an X-ray
observation that covers less than a full beat cycle, as was likely
the case here and for \pola.

Switching of the accretion between poles changes
the peaks in the power spectrum; this must be recognized if the periods
and accretion geometry are to be correctly identified.  
The spin period is often not present in the power spectrum if
there is pole switching, which can make the dominant peak
either $\omega-\Omega$ if $i+m$ is small, or split $\omega$
into the sidebands $\Omega$ and $2\omega-\Omega$ if $i+m$ is
large \citep{wyn92,wan20}.  In addition to the identifications of
$\omega$ and $\Omega$ possibly being reversed, an
ambiguity for an AP is whether to identify the upper sideband as
$\omega$ (model~1) or $2\omega-\Omega$ (model~2).
See, e.g., \citet{mey17}, \cite{lit19}, and \citet{wan20} on CD~Ind.
It was even inferred that accretion in CD~Ind and Paloma
switched from single- to two-pole around the beat cycle
\citep{hak19,lit22}, much like our model~1.

The alternative identifications of the main peak in the X-ray power
spectrum of \obj\ come with a beat period differing by a factor of 2,
which means that the observation either spans more than one beat
period (model~1) or less than one beat period (model~2),
with slightly different spin periods to match the pulse to the correct
pole.  Additional evidence that may bear on which model is correct
comes from the absorption dips.
The second pole seems to suffer photoelectric absorption for
up to three spin periods, which is most of the time that it is accreting
in model~1, but only a fraction of the time in model~2, where the second
pole eventually emerges looking much like the first pole in its pulse
profile.   It might seem that this is too much fine tuning for model~2.
On the other hand, the spacing of the absorption dips is more consistent
with the spin period of model~2 than model~1.  Furthermore, model~2
requires one out of the four dips to be coming from the opposite pole during
the other half of the beat cycle, which lends model~2 additional support.

A longer X-ray observation would likely resolve the ambiguity
of the beat frequency.  If model 1 is correct, a second 0.231~day cycle
will look much the same as the first, while in model 2 there should
be differences because a second pole will be accreting.
The uncertain outcome of the present study is due to the
limited length of the observation.  It is even possible that
the apparent change of state to two-pole accretion
(or flaring) is simply a random
event that will not repeat on a regular period.  In that case, the
simpler model~1 is probably correct as far as its period assignments
are concerned.  On the other hand, this supposed random event is what
is responsible for the strong peak in the hard X-ray power spectrum at
precisely the orbital frequency determined optically.
This would be quite a coincidence.

Whether or not there is pole switching also bears on the question of
whether there is an accretion disk, as hinted at by the optical
emission-line properties.  An X-ray signal at $2\omega-\Omega$
is a clear signature of stream-fed accretion.  In a conventional
IP that is fed from a truncated accretion disk, the inner disk has
erased all information about the location of the secondary.  So it is
difficult to imagine how a $2\omega$ spin signal could be modulated
strongly at the orbital frequency $\Omega$, which would require X-ray
reflection by material fixed in the orbital frame, while suppressing
the underlying spin signal.  If model 1 is correct {\it and\/} the apparent
beat-induced episode is not random, the same objection applies,
in the sense that the
inner accretion disk ``knows'' when to contribute to the
second pole.  Interestingly, the optical power spectra do have
signals involving the spin frequency, albeit
weaker than the ones at $\Omega$ and $2\Omega$.  These could be
from a contribution to the optical light from accretion columns,
or a reprocessing of the X-ray spin pulses on the secondary.

\section{Conclusions\label{sec:conc}}

  Using an archival \xmm\ observation, we identified candidates for the spin
  frequency of \obj\ that are either 24\% or 12\% faster than the orbital
  frequency determined optically.  We then refined their values using TESS and
  ATLAS photometry.  Either value of the frequency excess is characteristic
  of a small class of highly asynchronous polars or stream-fed IPs.
  The ambiguity arises because it is not clear if a complete pole-switching
  event took place during the observation.   If it did, it requires
  assigning $2\omega-\Omega$ to the main peak in the
  power spectrum instead of $\omega$ in the case of single-pole accretion.
  Absorption dips in the light curve tend to support $2\omega-\Omega$,
  therefore 12\% asynchronism.
  The most important follow-up to resolve this question would be a longer
  X-ray observation that covers at least a full cycle or two of the longer
  0.462~day beat candidate.  This should reveal whether a different pole is
  alternating in accretion with the first, through inevitable differences
  in viewing geometry, obscuration, or size.  A similar AP that has not
  yet been observed through its suggested 44~hr beat cycle
  is \pola\ \citep{hal17}.
  
  The optical emission lines of \obj\ do not display the properties
  common in ordinary polars, particularly the high velocities of the
  latter from radial accretion.  While radial velocities
  are generally expected to be modulated on the spin period if they come
  from deep in the accretion column, this appears not to be the case in
  several APs, including \obj, \pola, and \polb\
  \citep{tov17}, in which they represent the orbital period.  They may have
  to come from an accretion disk, or from extended streams that
  wrap around the WD to reach the magnetic poles.  The optical QPO
  of \obj\ is also a characteristic of disk systems.
  
  If there is a disk present, there must also be a stream that jumps
  over the disk to directly feed the magnetic poles, i.e.,
  the disk-overflow model of IPs \citep{hel93}.  Otherwise,
  the beating effect would be difficult to understand.  As to whether
  \obj\ is a pre-polar approaching synchronism, or an equilibrium IP,
  the magnetic accretion model of \citet{nor04} predicts that rotational
  equilibrium should only be possible when $P_{\rm s}/P_{\rm orb}<0.6$, which
  means that \obj\ should be in the processo of synchronizing.
  Indeed, the candidates for
  nearly synchronous IPs listed in \citet{nor04} with $P_{\rm s}/P_{\rm orb}>0.7$
  have subsequently been judged instead to be already synchronous polars\footnote{https://asd.gsfc.nasa.gov/Koji.Mukai/iphome/iphome.html}, with the exception
  of Paloma, the AP that \obj\ most closely resembles.

  \begin{acknowledgements}
    I thank the anonymous referee for recommending several improvements,
    and offering generous advice, including finding evidence in TESS and
    ATLAS data. Slavko Bogdanov provided assistance with the \xmm\ OM data.
    John Thorstensen and Colin Littlefield obliged me with helpful discussions.
    Joe Patterson warned me years ago that the 975~s signal is probably a QPO.

This work was supported by NASA grant 80NSSC21K0819, and is based on
observations obtained with XMM-Newton,
an ESA science mission with instruments and contributions directly funded
by ESA Member States and NASA.  This paper includes data collected with the
TESS mission, obtained from the MAST data archive at the Space Telescope
Science Institute (STScI). Funding for the TESS mission is provided by the
NASA Explorer Program.  The ATLAS project is primarily funded
to search for near earth asteroids through NASA grants NN12AR55G, 80NSSC18K0284,
and 80NSSC18K1575. The ATLAS science products have been made possible through
the contributions of the University of Hawaii Institute for Astronomy,
the Queen's University Belfast, the Space Telescope Science Institute,
the South African Astronomical Observatory, and The Millennium Institute
of Astrophysics (MAS), Chile.
\end{acknowledgements}
  
\facility{XMM, MDM, TESS, ATLAS}

\end{document}